\documentclass[11pt]{article}
\begin{document}
\pagestyle{plain}
\setcounter{page}{1}
\begin{center}
{\large\bf Quantum Gravity Solution To The Cosmological Constant
Problem}
\vskip 0.1 true in {\large J. W. Moffat}\footnote{e-mail:
john.moffat@utoronto.ca}  \vskip 0.1 true in {\it Department of
Physics, University of Toronto, Toronto, Ontario M5S 1A7,
Canada} \vskip 0.1 true in and \vskip 0.1 true in {\it Perimeter
Institute for Theoretical Physics, Waterloo, Ontario N2J 2W9,
Canada}
\date{\today}
\begin{abstract}%
A nonlocal quantum gravity theory is presented which is finite and unitary
to all orders of perturbation theory. Vertex form factors in Feynman
diagrams involving gravitons suppress graviton and matter vacuum
fluctuation loops by introducing a low-energy gravitational scale,
$\Lambda_{\rm Gvac} < 2.4\times 10^{-3}$ eV. Gravitons coupled to
non-vacuum matter loops and matter tree graphs are controlled by a vertex
form factor with the energy scale, $\Lambda_{GM}< 1-10$ TeV.
\end{abstract}

\end{center}
\vskip 0.1 true in
Talk given at the XVIIIth IAP Colloquium: Observational and
Theoretical Results on the Accelerating Universe, July 1-5, 2002,
Paris, France.
\vskip 0.1 true in

\section{\bf Gravitational Coupling to Vacuum Energy}

We can define an effective cosmological constant~\cite{Straumann}.

\begin{equation}
\lambda_{\rm eff}=\lambda_0+\lambda_{\rm vac},
\end{equation}
where $\lambda_0$ is the ``bare'' cosmological
constant in Einstein's classical field equations,
and $\lambda_{\rm vac}$ is the contribution that arises from the
vacuum density $\lambda_{\rm vac}=8\pi G\rho_{\rm vac}$.

Already at the standard model electroweak scale $\sim 10^2$ GeV, a
calculation of the vacuum density $\rho_{\rm vac}$, based on local quantum field
theory, results in a discrepancy of order $10^{55}$ with the observational
bound
\begin{equation}
\label{vacbound}
\rho_{\rm vac} \leq 10^{-47}\, ({\rm GeV})^4.
\end{equation}
This results in a severe fine-tuning problem of order $10^{55}$,
since the virtual quantum fluctuations giving rise to $\lambda_{\rm vac}$
must cancel $\lambda_0$ to an unbelievable degree of accuracy.
This is the ``particle physics'' source of the cosmological
constant problem.

\section{\bf Nonlocal Quantum Gravity}

Let us consider a model of nonlocal gravity with the action
$S=S_g+S_M$, where ($\kappa^2=32\pi G$)
\footnote{The present version of a nonlocal quantum gravity and
field theory model differs in detail from
earlier published work~\cite{Moffat,Moffat2}. A paper is in preparation in which more complete
details of the model will be provided.}:
\begin{equation}
S_g=-\frac{2}{\kappa^2}\int d^4x\sqrt{-g}\biggl\{R[g,{\cal
G}^{-1}]+2\lambda_0\biggr\}
\end{equation}
and $S_M$ is the matter action, which for the simple case of
a scalar field $\phi$ is given by
\begin{equation}
S_M=\frac{1}{2}\int d^4x\sqrt{-g}{\cal
G}^{-1}\biggl(g^{\mu\nu}\nabla_\mu\phi{\cal F}^{-1}\nabla_\nu\phi
-m^2\phi{\cal F}^{-1}\phi\biggr).
\end{equation}
Here, ${\cal G}$ and ${\cal F}$ are nonlocal regularizing,
{\it entire} functions and $\nabla_\mu$ is the covariant
derivative with respect to the metric $g_{\mu\nu}$. As an example,
we can choose the covariant
functions
${\cal G}(x)=\exp\biggl[-{\cal D}(x)/\Lambda_G^2\biggr]$,
and
${\cal F}(x)=\exp\biggl[-({\cal D}(x)+m^2)/\Lambda_M^2\biggr]$,
where ${\cal D}\equiv\nabla_\mu\nabla^\mu$, and $\Lambda_G$ and
$\Lambda_M$ are gravitational and matter energy scales,
respectively~\cite{Moffat,Moffat2}.

We expand $g_{\mu\nu}$ about flat Minkowski spacetime:
$g_{\mu\nu}=\eta_{\mu\nu}+\kappa h_{\mu\nu}$. The propagators for the
graviton and the $\phi$ field in a fixed gauge are given by
\begin{equation}
{\bar D}^\phi(p)=\frac{{\cal G}(p){\cal F}(p)}{p^2-m^2+i\epsilon},
\end{equation}
$$ $$
\begin{equation}
{\bar D}^G_{\mu\nu\rho\sigma}(p)=\frac{(\eta_{\mu\rho}\eta_{\nu\sigma}
+\eta_{\mu\sigma}\eta_{\nu\rho}-\eta_{\mu\nu}\eta_{\rho\sigma})
{\cal G}(p)}{p^2+i\epsilon}.
\end{equation}

Unitarity is maintained for the S-matrix, because ${\cal G}$ and
${\cal F}$ are {\it entire} functions of $p^2$, preserving the Cutkosky
rules. Gauge invariance can be maintained by satisfying certain
constraint equations for ${\cal G}$ and ${\cal F}$ in every order of
perturbation theory.  This guarantees that $\nabla_\nu T^{\mu\nu}=0$.

\section{\bf Resolution of the CCP}

Due to the equivalence principle {\it gravity couples to all
forms of energy}, including the vacuum energy density $\rho_{\rm
vac}$, so we cannot ignore these virtual quantum
fluctuations in the presence of a non-zero gravitational field.
Quantum corrections to $\lambda_0$ come from loops formed from
massive standard model (SM) states, coupled to external graviton
lines at essentially zero momentum.

Consider the dominant contributions to the vacuum
density arising from the graviton-standard model loop corrections.
We shall adopt a model consisting of a photon loop coupled to
gravitons, which will contribute to the vacuum polarization loop coorection
to the bare cosmological constant $\lambda_0$. The covariant photon action
is~\cite{Leibbrandt}:
\begin{equation}
S_A=-\frac{1}{4}\sqrt{-g}g^{\mu\nu}g^{\alpha\beta}{\cal
G}^{-1}F_{\mu\alpha}{\cal F}^{-1}F_{\nu\beta},
\end{equation} with
$F_{\mu\alpha}=\partial_\mu A_\alpha-\partial_\alpha A_\mu$.

The lowest order correction to the graviton-photon vacuum loop will have
the form (in Euclidean momentum space):
\begin{equation}
\Pi^{\rm Gvac}_{\mu\nu\rho\sigma}(p)
=\kappa^2\int\frac{d^4q}{(2\pi)^4}V_{\mu\nu\lambda\alpha}(p,-q,-q-p)
$$ $$
\times{\cal
F}^\gamma(q^2)D^\gamma_{\lambda\beta}(q^2)V_{\rho\sigma\beta\gamma}(-p,q,p-q)
{\cal F}^\gamma((p-q)^2)D^\gamma_{\alpha\gamma}((p-q)^2){\cal G}^{\rm
Gvac}(q^2), \end{equation}
where $V_{\mu\nu\rho\sigma}$ is the
photon-photon-graviton vertex and in a fixed gauge:
$D^\gamma_{\mu\nu}=-\delta_{\mu\nu}/q^2$
is the free photon propagator. Additional contributions to
$\Pi^{\rm Gvac}_{\mu\nu\rho\sigma}$ come from tadpole
graphs~\cite{Leibbrandt}.

This leads to the vacuum polarization tensor
\begin{equation}
\label{Ptensor}
\Pi^{\rm Gvac}_{\mu\nu\rho\sigma}(p)=\kappa^2
\int\frac{d^4q}{(2\pi)^4}\frac{1}{q^2[(q-p)^2]}
$$ $$
\times K_{\mu\nu\rho\sigma}(p,q)
\exp\biggl\{-q^2/\Lambda^2_M
-[(q-p)^2]/\Lambda^2_M-q^2/\Lambda^2_{{\rm Gvac}}\biggr\}.
\end{equation}
For $\Lambda_{{\rm Gvac}} \ll \Lambda_M$, we
observe that from power counting of the momenta in the loop integral, we
get
\begin{equation}
\Pi^{\rm {\rm Gvac}}_{\mu\nu\rho\sigma}(p)\sim
\kappa^2\Lambda_{{\rm Gvac}}^4N_{\mu\nu\rho\sigma}(p^2)
\sim\frac{\Lambda_{{\rm Gvac}}^4}{M^2_{\rm PL}}N_{\mu\nu\rho\sigma}(p^2),
\end{equation}
where $N(p^2)$ is a finite remaining part of $\Pi^{\rm {\rm Gvac}}(p)$ and
$M_{\rm PL}\sim 10^{19}$ GeV is the Planck mass.

We now have
\begin{equation}
\rho_{\rm vac}\sim M^2_{\rm PL}\Pi^{\rm {\rm Gvac}}(p)\sim\Lambda_{{\rm
Gvac}}^4.
\end{equation} If we choose $\Lambda_{{\rm Gvac}}\leq 10^{-3}$
eV, then the quantum correction to the bare cosmological constant
$\lambda_0$ is suppressed sufficiently to satisfy the observational bound
on $\lambda$, {\it and it is protected from large unstable radiative
corrections}.

This provides a solution to the
cosmological constant problem at the energy level of the standard model
and possible higher energy extensions of the standard model. The universal
fixed gravitational scale $\Lambda_{{\rm Gvac}}$ corresponds to the fundamental
length $\ell_{{\rm Gvac}}\leq 1$ mm at which virtual gravitational radiative
corrections to the vacuum energy are cut off.

The gravitational form factor ${\cal G}$, {\it when
coupled to non-vacuum SM gauge boson or matter loops}, will have the form
in Euclidean momentum space
${\cal G}^{\rm GM}(q^2)
=\exp\biggl[-q^2/\Lambda_{GM}^2\biggr]$.
If we choose $\Lambda_{GM} = \Lambda_{M}> 1-10$ TeV, then we will
reproduce the standard model experimental results, including the running
of the standard model coupling constants, and ${\cal G}^{GM}(q^2)={\cal
F}^M(q^2)$ becomes ${\cal G}^{GM}(0)={\cal F}^M(q^2=m^2)=1$ on the mass
shell. {\it This solution to the CCP leads to a violation of the WEP for
coupling of gravitons to vacuum energy and matter.} This
could be checked experimentally in a satellite E\"otv\"os experiment on the
Casimir vacuum energy.

We observe that the required suppression of the vacuum diagram
loop contribution to the cosmological constant, associated with
the vacuum energy momentum tensor at lowest order,
demands a low gravitational energy scale $\Lambda_{{\rm Gvac}}\leq
10^{-3}$ eV, which controls the coupling of gravitons to
pure vacuum graviton and matter fluctuation loops.

In our finite, perturbative
quantum gravity theory nonlocal gravity produces a long-distance
infrared cut-off of the vacuum energy density through the low energy
scale $\Lambda_{{\rm Gvac}} < 10^{-3}$ eV~\cite{Moffat2}.
Gravitons coupled to {\it non-vacuum} matter tree graphs and matter loops
are controlled by the energy scale: $\Lambda_{GM}=\Lambda_{M} > 1-20$
TeV

The rule is: When external graviton lines are removed from a
matter loop, leaving behind {\it pure} matter fluctuation vacuum loops,
then those initial graviton-vacuum loops are suppressed by the form factor
${\cal G}^{{\rm Gvac}}(q^2)$ where $q$ is the internal matter loop momentum and
${\cal G}^{{\rm Gvac}}(q^2)$ is controlled by $\Lambda_{{\rm Gvac}} \leq 10^{-3}$ eV.
On the other hand, e.g. the proton first-order self-energy graph, coupled
to a graviton is controlled by $\Lambda_{GM}=\Lambda_M > 1-20$ TeV {\it
and does not lead to a measurable violation of the equivalence principle.}

The scales $\Lambda_M$ and $\Lambda_{{\rm Gvac}}$ are determined in
loop diagrams by the quantum non-localizable nature of the gravitons and
standard model particles. The gravitons coupled to matter and matter
loops have a nonlocal scale at $\Lambda_{GM}=\Lambda_M > 1-20$ TeV or a
length scale $\ell_M < 10^{-16}$ cm, whereas the gravitons coupled to
pure vacuum energy are localizable up to an energy scale
$\Lambda_{{\rm Gvac}}\sim 10^{-3}$ eV or down to a length scale $\ell_{{\rm Gvac}} > 1$
mm.

The fundamental energy scales $\Lambda_{{\rm Gvac}}$ and
$\Lambda_{GM}=\Lambda_M$ are determined by the underlying physical nature
of the particles and fields and do not correspond to arbitrary cut-offs,
which destroy the gauge invariance, Lorentz invariance and unitarity of the
quantum gravity theory for energies $>\Lambda_{{\rm Gvac}}\sim 10^{-3}$ eV.
The underlying explanation of these physical scales must be sought in a
more fundamental theory\footnote{It is interesting to note that if we
choose $\Lambda_{\rm GM}=\Lambda_M=5$ TeV, then we obtain
$\Lambda_{\rm Gvac}=\Lambda_M^2/M_{\rm PL}=2.1\times 10^{-3}$
eV.}

Let us consider the cosmological problem in the context of our
results. For a spatially flat universe, the Friedmann equation
is
$\label{Friedmann}
H^2\sim 8\pi G{\bar\rho}_{\rm vac}/3$,
where $H={\dot R}/R$, $R$ denoting the cosmic
scale and ${\bar\rho}_{\rm vac}=\lambda_0/8\pi
G+\rho_{\rm vac}$ is the effective vacuum energy density,
including the bare vacuum energy density contribution, and we
have assumed that the vacuum energy dominates the Friedmann
equation. If we assume that the vacuum energy density has its
``natural'' value with a cutoff $\Lambda_c\sim  M_{PL}$, then the
universe never expands beyond a ridiculously small size.
However, since we have suppressed the vacuum energy density
contribution to the Friedmann equation by our nonlocal quantum
gravity calculation, we will find that
$H < H_0\sim 10^{-33}$ eV in agreement with observations.

Our calculations are based on perturbation theory, so we cannot
address possible non-perturbative contributions to the vacuum
density, such as those arising from gluon condensates or
spontaneous symmetry breaking phase transitions. These
contributions will have to be suppressed in a non-perturbative
nonlocal quantum gravity calculation.

\vskip 0.1 true in
{\bf Acknowledgments}
\vskip 0.1 true in
I thank Michael Clayton for helpful and
stimulating discussions. This work was supported by the Natural Sciences and
Engineering Research Council of Canada.  \vskip 0.1 true in


\begin{thebibliography}{100}

\bibitem{Straumann} Weinberg, S., Rev. Mod. Phys. {\bf 61}, 1
(1989); Straumann, N., arXive:astro-ph/0203330; Peebles, P. J.
E., and Ratra, B., arXive:astro-ph/0207347; Ellwanger, U., arXive:hep-ph/0203252.
\bibitem{Moffat} Moffat, J. W.,
Phys. Rev. {\bf D41}, 1177 (1990); Evens, D., Moffat, J. W.,
Kleppe, G., and Woodard, R. P., Phys. Rev. {\bf D43}, 499
(1991).  \bibitem{Moffat2} Moffat, J. W., arXive:hep-ph/0102088
v2.  \bibitem{Leibbrandt} Capper, D. M., Leibbrandt, G.,
Medrano, M. R., Phys. Rev. {\bf D8}, 4320 (1973); Capper, D. M.,
Duff, M. J., and Halpern, L., Phys. Rev. {\bf 10}, 461 (1974).

\end{thebibliography}
\end{document}